\documentclass[submission,copyright,creativecommons]{eptcs}
\usepackage{amsmath,amssymb,amsfonts,amsthm}
\usepackage{fancyvrb}
\usepackage[shortcuts]{extdash}
\usepackage{color}

\usepackage{listings}
\newcommand*{\var}[1]{\mathit{#1}}

\title{Term-Level Reasoning in Support of Bit-blasting}
\author{Sol Swords
  \institute{Centaur Techology, Inc.\\
    Austin, TX, USA}
  \email{sswords@centtech.com}
}
\def\titlerunning{Term-Level Reasoning in Support of Bit-Blasting}
\def\authorrunning{S. Swords}
\begin{document}
\maketitle

\begin{abstract}

  GL is a verified tool for proving ACL2 theorems using Boolean
  methods such as BDD reasoning and satisfiability checking.  In its
  typical operation, GL recursively traverses a term, computing a
  symbolic object representing the value of each subterm.  In older
  versions of GL, such a symbolic object could use Boolean functions
  to compactly represent many possible values for integer and Boolean
  subfields, but otherwise needed to reflect the concrete structure of
  all possible values that its term might take.  When a term
  has many possible values that can't share such a
  representation, this can easily cause blowups because GL must then
  case-split.  To address this problem, we have added several features
  to GL that allow it to reason about term-like symbolic objects using
  various forms of rewriting.  These features allow GL to be
  programmed with rules much like the ACL2 rewriter, so that users may
  choose a better normal form for terms for which the default,
  value-like representation would otherwise cause case explosions.  In
  this paper we describe these new features; as a motivating example,
  we show how to program the rewriter to reason effectively about the
  theory of records.

\end{abstract}

\section{Introduction}
\label{sec:intro}

GL is an ACL2 framework for proving theorems using bit-level methods,
namely BDD reasoning or SAT solving \cite{bit-blasting-GL, gl-diss}.
It works by computing bit-level symbolic representations of terms,
which we also describe as symbolically executing them.  To prove a
theorem, it computes such a representation of the theorem's conclusion
and thus reduces it to a satisfiability check: if the negation of the Boolean formula
representing the conclusion is unsatisfiable, then the theorem is
proved.

This paper introduces term-level reasoning capabilities that are now
built into GL.  These capabilities allow GL to prove theorems at a
higher level of abstraction, without computing a concrete
representation for the value of each subterm encountered.  It also
allows theorems to be proved about unconstrained variables by
generating bit-level representations for accessors applied to those
variables.  We have verified the soundness of all of these extensions
in order to admit GL as a verified clause processor.

\subsection{Extensions to ``Traditional'' GL}

GL has always contained a symbolic object
representation for function calls and unconstrained variables, but
previously the use of such objects would almost guarantee failure.
GL's previous use was {\em value-oriented}, as indicated by the use of the
term ``symbolic execution'': the GL system would crawl over a term
just as a simple applicative Lisp interpreter would, producing a
(symbolic) value for each subterm.  Typical symbolic values expected
to appear during a successful symbolic execution could include:

\begin{itemize}
\item symbolic Booleans -- Boolean functions in a BDD or
  AIG representation, depending on the mode of operation
\item symbolic integers -- represented as a list of Boolean functions,
  one for each two's-complement bit of the integer
\item concrete values -- represented by themselves (except for objects
  containing certain tagging symbols used to delineate kinds of
  symbolic objects, which could then be tagged to ``escape'' them)
\item conses of symbolic values.
\end{itemize}

In certain cases the symbolic values might also be expected to contain:

\begin{itemize}
\item if-then-else constructs, so that a symbolic value could take
  various different concrete values or different types of symbolic
  value depending on symbolic conditions.
\end{itemize}

But two other kinds of symbolic objects existed in the representation
primarily for use in bad cases:

\begin{itemize}
\item function call -- representing a function applied to a list of
  symbolic values, primarily used to indicate that GL was unable to
  compute a useful representation for some subterm
\item variable -- representing a free variable of unknown type, used
  when a theorem variable lacked a specific symbolic representation
  provided by the user.
\end{itemize}

The new GL features described in this paper allow it to operate in a
{\em term-oriented} manner, where the above function and variable
representations can also be manipulated productively.  We describe the
new features supporting this in Section~\ref{sec:features} and discuss
some implementation details that may be relevant to users in
Section~\ref{sec:details}.
We show a complete set of rules for
term-level manipulation of records in Section~\ref{sec:example}, then
describe how this new feature set has been applied to a class of
industrial proofs in Section~\ref{sec:applications}.   We discuss some issues involved in the proof 
of soundness of these new features in Section~\ref{sec:proof}.  Some possible
future extensions are discussed in Section~\ref{sec:future}.

\section{Term-level Features of GL}
\label{sec:features}
The term-level capabilities of GL consist of the following new
features. 
While complete documentation of these features is beyond
the scope of this paper (see the ACL2 community books
documentation~\cite{acl2:doc}, including topic
\href{http://www.cs.utexas.edu/users/moore/acl2/manuals/current/manual/?topic=GL____TERM-LEVEL-REASONING}{\underline{gl::term-level-reasoning}}), we motivate and briefly describe each of them in this section:

\begin{itemize}
\item Conditional rewriting with user-defined rules integrated with the symbolic interpreter (Section~\ref{sec:rewriting})
\item Generation and tracking of fresh Boolean variables representing truth values of term-level objects (Section~\ref{sec:vargen})
\item Rules to generate counterexamples from Boolean valuations for terms (Section~\ref{sec:counterexes})
\item Rules to add constraints among the generated Boolean variables (Section~\ref{sec:constraintgen})
\item Rules that merge \texttt{if} branches containing different term-level objects that could be represented in a common way (Section~\ref{sec:branchmerge})
\item Support for binding theorem variables to shape specifiers representing function calls (Section~\ref{sec:callshapespec}).
\end{itemize}

\subsection{Abstracting Implementation Details with Rewriting}
\label{sec:rewriting}
Suppose we are using the popular records library found in the
community books file ``misc/records.lisp''
\cite{kaufmann2002efficient}, and we have the term \texttt{(s :a b nil)}.
Supposing $\var{b}$ is a 4-bit integer, and we have its symbolic
representation as a vector of 4 Boolean functions, we can represent
this call as \texttt{'((:a . $\var{b}$))}.
But suppose instead that $\var{b}$ is a Boolean.  The default value
of any key in a record is \texttt{nil}, so the \texttt{s} function
does not add pairs to record structures when binding them to
\texttt{nil}.  So the best we can do to represent this value is a term-like representation,
\begin{lstlisting}
  (if <$\var{b}$>  '((:a . t)) 'nil)
\end{lstlisting}
This effectively introduces a case-split: every further operation on
this record must consider separately the case where $\var{b}$ is
\texttt{t} versus \texttt{nil}.  We can easily get an exponential
blowup in our representation by setting several keys to symbolic
Boolean values.  This is especially annoying since the elision of
pairs that bind keys to \texttt{NIL} is just an implementation detail
of the library, not particularly relevant to the meaning of what is
happening.

A better way to deal with this is to avoid representing the record as
a cons structure.  Instead, GL can treat the record library's
\texttt{s} and \texttt{g} as uninterpreted, and use rewrite rules to
resolve them.  The following forms declare these functions uninterpreted:
\begin{verbatim}
(gl::gl-set-uninterpreted g)
(gl::gl-set-uninterpreted s)
\end{verbatim}
When the symbolic interpreter reaches a call of \texttt{g} or
\texttt{s}, it will compute symbolic representations for the arguments
to the call.  Normally, it would then symbolically interpret the
definition of the function.  These declarations say to instead simply
return an object representing the call of \texttt{g} or \texttt{s}
applied to those symbolic arguments.  These function call objects can
then be operated on by rewrite rules.

GL's rewrite rules may be declared using \texttt{gl::def-gl-rewrite};
this stores the theorem as an ACL2 rewrite rule but disables it and
adds it to a table that determines which rules GL will try.  For
records, some proofs can be completed with only the case-splitting
version of the simple get-of-set rule:
\begin{verbatim}
(gl::def-gl-rewrite g-of-s-for-gl
 (equal (g k1 (s k2 v r))
        (if (equal k1 k2)
            v
          (g k1 r))))
\end{verbatim}
\noindent In general, a more comprehensive set of rules is required;
we discuss these in Section \ref{sec:example}.  With this rule in
place, if a \texttt{g} call is encountered and the symbolic
representation of its second argument is a call of \texttt{s}, then
this rule applies, and instead of producing a \texttt{g} call object,
the symbolic interpreter is instead applied to the right-hand side of
this rule with the appropriate substitution.

GL's rewriter supports a subset of the ACL2 rewriter's features.  It
can use rules that rewrite on \texttt{equal} or \texttt{iff}
equivalences.  It supports conditional rewriting by backchaining to
relieve hypotheses.  It supports \texttt{syntaxp} hypotheses, but
these aren't compatible with ACL2's syntaxp hypotheses, because ACL2's
use term syntax whereas GL's use symbolic object syntax.  Some further
improvements may be considered in future work, which we discuss in Section
\ref{sec:parity}.

\subsection{Boolean Representations for Unconstrained Variables and Accessors}
\label{sec:vargen}
Traditionally, most theorems proved using GL have had hypotheses
constraining each variable to a type that could be represented using a
bounded number of bits, such as \texttt{(unsigned-byte-p 8 x)}.  But
in some cases this shouldn't be necessary because only certain bits
are accessed from the variable regardless of its size or type.
Consider:
\begin{verbatim}
(equal (lognot (lognot (loghead 5 x))) (loghead 5 x))
\end{verbatim}
This is true for any value of \texttt{x}, even non-integers; but
previously, GL would not be able to prove this without constraining
\texttt{x} to be a member of a bounded type.  However, GL's new
term-level capabilities include the ability to generate fresh Boolean
variables to represent certain terms.  Specifically, GL generates a
fresh Boolean variable for any \texttt{if} test that cannot be
otherwise resolved to a symbolic Boolean value.  Therefore, a rewrite
rule such as the following may be used to introduce new Boolean
variables for \texttt{loghead} calls such as those above.  (Note that the
syntaxp test in this rule applies to the GL symbolic object
representation, not the ACL2 term representation!)
\begin{verbatim}
(gl::def-gl-rewrite expand-loghead-bits
  (implies (syntaxp (integerp n))
           (equal (loghead n x)
                  (if (zp n)
                      0
                    (logcons (if (logbitp 0 x) 1 0)
                             (loghead (1- n) (ash x -1)))))))
\end{verbatim}
This rewrite rule unwinds the \texttt{loghead} call of the theorem
above.  Each \texttt{logbitp} call first simply produces a term-level
representation, but then, since the term is used as an \texttt{if} test, a new Boolean variable is generated to represent it.  In particular, Boolean variables
are introduced for the following terms:
\begin{verbatim}
(logbitp 0 x)
(logbitp 0 (ash x -1))
(logbitp 0 (ash (ash x -1) -1)) ...
\end{verbatim}
We can tweak this representation via the following rewrite rule to a more uniform \texttt{(logbitp $n$ x)}:
\begin{verbatim}
(gl::def-gl-rewrite logbitp-of-ash-minus1
  (implies (syntaxp (integerp n))
           (equal (logbitp n (ash x -1))
                  (logbitp (+ 1 (nfix n)) x))))
\end{verbatim}

All of the \texttt{logbitp} test terms are generated twice, once for each call
\texttt{(loghead 5 x)}, but the second time each one is recognized and
the Boolean variable generated the first time is returned instead of
generating a fresh one.

This strategy can be used to reason about more complex structures.
For example, if a conjecture references an arbitrary record but only
accesses a fixed set of fields and a fixed number of bits of each
field, then it can generate Boolean variables as above for the fields,
such as \texttt{(logbitp $n$ (g $\var{field}$ x))}.

\subsection{Generating Term-level Counterexamples from Boolean Counterexamples}
\label{sec:counterexes}

GL's method of proving a theorem is to obtain a Boolean formula 
representing the conclusion and then ask
whether it is valid, either by examination of the canonical BDD
representation or by querying a SAT solver to determine whether the
negation of the AIG representation is satisfiable.  When the result is valid (its
negation is unsatisfiable), this proves the theorem, at least 
when all side obligations are met.
But when it is not valid, a contradiction at the Boolean
level does not directly yield a real counterexample to the theorem ---
instead, it gives a Boolean valuation to each of the terms that were
assigned Boolean variables as described above.\footnote{This problem
  does not exist in traditional GL, where all variables of a
  conjecture are given bit-level symbolic representations by the
  g-bindings, and no other Boolean variables are generated.  A
  bit-level counterexample then directly yields valuations for the
  term variables by evaluating their symbolic representations under
  the counterexample assignment.}
Consider the following example:
\begin{verbatim}
(gl::def-gl-thm minus-logext-minus-loghead-is-logext-loghead
   :hyp t
   :concl (equal (- (logext 5 (- (loghead 5 x))))
                 (logext 5 (loghead 5 x)))
   :g-bindings nil
   :rule-classes nil)
\end{verbatim}
For this conjecture, a counterexample produced by the SAT solver
includes the following assignments for the \texttt{logbitp} terms that
make up the \texttt{loghead} (as described in the previous section):
\begin{verbatim}
(logbitp 0 x) = NIL
(logbitp 1 x) = NIL
(logbitp 2 x) = NIL
(logbitp 3 x) = NIL
(logbitp 4 x) = T
\end{verbatim}
We would
like to translate the valuations for these terms into a value for \texttt{x} itself.  In this
case, it seems obvious that we should assign \texttt{x} the value 16,
or -16, or any integer whose low 5 two's-complement bits are 16: there isn't a
unique assignment satisfying the given Boolean valuation.
Additionally, it may or may not even be possible to set the term
variables such that the terms all have their assigned values.  For
example, a badly configured set of rewrite rules could result in
\texttt{(logbitp 0 (ash x -1))} and \texttt{(logbitp 1 x)} being
assigned independent Boolean values; if those values differ, then
there is no \texttt{x} for which both valuations hold.  Furthermore, since
Boolean variables can be assigned to arbitrary terms, it could be an
arbitrarily hard problem to simultaneously satisfy the Boolean
valuations of the terms.  So we are looking for a heuristic solution
that works in practice, since a total solution doesn't exist.

Our solution is to allow the user to specify rewrite rules that
describe how to update variables and subterms in response to
assignments.  Every variable starts out assigned to \texttt{NIL}, and
assignments to terms involving the variables may cause that value to
be updated if appropriate rules are in place.  With no such rules, the
above theorem fails to prove but the counterexample produced assigns
\texttt{NIL} to \texttt{x}, which is a false counterexample.  However,
if we add the following rule, then it will instead assign 16 to
\texttt{x}, which is a real counterexample:
\begin{verbatim}
(gl::def-glcp-ctrex-rewrite
  ((logbitp n i) v)
  (i (bitops::install-bit n (bool->bit v) i)))
\end{verbatim}
This rule reads: ``If a term matching \texttt{(logbitp n i)} is
assigned value \texttt{v}, then instead assign \texttt{i} the
evaluation of \texttt{(bitops::install-bit n (bool->bit v) i)}.''  With
this rule in place GL correctly generates the counterexample \texttt{x
  = 16} for the above proof attempt.  These rules do not incur any
proof obligations, since they are not used for anything relevant to
soundness.  Note also that these operations are performed without 
guard checking, so that installing a bit into \texttt{NIL} does not 
cause an error.

Counterexample rewrite rules work together to resolve nested accesses.
For example, we could have another rule for dealing with record accesses:
\begin{verbatim}
(gl::def-glcp-ctrex-rewrite
  ((g field rec) v)
  (rec (s field v rec)))
\end{verbatim}
\noindent This rule can work in concert with the \texttt{logbitp} rule
above to resolve assignments to terms such as \texttt{(logbitp 4 (g
  :fld x)) = t}.  Suppose that before processing this assignment, \texttt{x} was bound to \texttt{nil}.
The \texttt{logbitp} rule fires
first and replaces our assignment with \texttt{(g :fld x) =
  (bitops::install-bit 4 (bool->bit t) (g :fld nil))}, which evaluates
to 16.  Then the rule for \texttt{g} applies, and results in the
assignment \texttt{x = (s :fld 16 nil)}, which evaluates to
\texttt{((:FLD . 16))}.  This is then the provisional value for
\texttt{x}, until it is updated by the resolution of another Boolean assignment.

Most counterexample rewrite rules essentially describe how to update
an aggregate so that a particular field of that aggregate will have a
given value: the \texttt{logbitp} rule above shows how to update a
particular bit of an integer, and the \texttt{g} rule shows how to
update a field in a record.

\subsection{Adding Constraints among Generated Boolean Variables}
\label{sec:constraintgen}
When assigning Boolean variables to arbitrary term-level conditions, a
pitfall is that by default all such variables are assumed to be
independent.  As noted in Section~\ref{sec:counterexes}, this isn't
guaranteed; it may be impossible for the Boolean valuation to be
satisfied by an assignment to the term-level variables.  The following
form shows a proof attempt that fails due to a Boolean assignment that can't be realized by a term-level assignment:
\begin{verbatim}
(gl::def-gl-thm loghead-equal-12-when-integerp
  :hyp t
  :concl (equal (and (integerp x)
                     (equal (loghead 5 x) 12))
                (equal (loghead 5 x) 12))
  :g-bindings nil)
\end{verbatim}
\noindent The Boolean formula given to the SAT solver for this
conjecture has variables for \texttt{(integerp x)} and for
\texttt{(logbitp $\var{n}$ x)} for $\var{n}$ from 0 to 4.  The
counterexample it returns has \texttt{(integerp x) = NIL} and the
\texttt{logbitp}s set so that \texttt{(loghead 5 x) = 12}.  But this
is impossible because, e.g., \texttt{(logbitp 2 x) = NIL} if
\texttt{x} is not an integer: in fact, \texttt{(logbitp $\var{n}$
  x)} implies \texttt{(integerp x)}.  GL allows a kind of rule that
encodes these sorts of constraints among Boolean conditions:
\begin{verbatim}
(gl::def-gl-boolean-constraint logbitp-implies-integerp
  :bindings ((bit0 (logbitp n x))
             (intp (integerp x)))
  :body (implies bit0 intp))
\end{verbatim}
\noindent This form submits the following theorem:
\begin{verbatim}
(let ((bit0 (if (logbitp n x) t nil))
      (intp (if (integerp x) t nil)))
  (implies bit0 intp))
\end{verbatim}
\noindent It additionally adds information about the constraint rule
to a database that is used to track relationships among GL's generated
Boolean variables.  In this case it keeps track of Boolean variables
added for terms matching \texttt{(logbitp n x)} and \texttt{(integerp
  x)}, and for each combination of these terms it evaluates the
theorem body and adds the resulting Boolean formula to a list of
constraints.  When the theorem's final SAT query is constructed, all
such constraints are included in the formula to be checked, preventing
the false counterexamples that led to our proof failure above.  With
this constraint rule, the theorem proves.

The distinction between bindings and body in a Boolean constraint rule
has important effects on how the rule is applied.  All bindings must be
matched to terms with existing Boolean variables before the rule applies.  So in the version
above, both \texttt{(logbitp n x)} and \texttt{(integerp x)} terms
must be encountered as \texttt{if} tests before the rule fires.
We can rewrite the rule as follows so that the constraint is added as
soon as a term matching \texttt{(logbitp n x)} is encountered:
\begin{verbatim}
(gl::def-gl-boolean-constraint logbitp-implies-integerp-stronger
  :bindings ((bit0 (logbitp n x)))
  :body (implies bit0 (integerp x)))
\end{verbatim}

Constraint rules can be used to perform a kind of forward reasoning.
Suppose that we used terms such as
\begin{verbatim}
(logbitp 0 (ash (ash (... (ash x -1) ...) -1) -1))
\end{verbatim}
\noindent as our normal form for bits extracted from a variable
\texttt{x}, rather than \texttt{(logbitp $n$ x)}; this occurs if we
omit the rule \texttt{logbitp-of-ash-minus1} stated in Section~\ref{sec:vargen}.  The
following two rules then suffice to prove our theorem:
\begin{verbatim}
(gl::def-gl-boolean-constraint logbitp-implies-nonzero
  :bindings ((bit0 (logbitp n i)))
  :body (implies bit0 (and (integerp i) (not (equal 0 i)))))

(gl::def-gl-boolean-constraint nonzero-rsh-implies-nonzero
  :bindings ((iszero (equal 0 (ash i -1))))
  :body (implies (not iszero) (and (integerp i) (not (equal 0 i)))))
\end{verbatim}
\noindent Whenever a \texttt{logbitp} term is derived from the
\texttt{loghead} in our theorem, this triggers the first rule.  The
body of that rule is evaluated in an \texttt{if}-test-like context, so
that \texttt{(integerp i)} and \texttt{(equal 0 i)} both are assigned
Boolean variables if necessary.  Then if \texttt{i} matched a nesting
of \texttt{ash} terms, the addition of \texttt{(equal 0 i)} will lead
to the second rule firing repeatedly until \texttt{i} unifies with the
innermost variable.  Together these set up a chain of implications so
that ultimately, if the \texttt{logbitp} is true, it is known that the
inner variable is an integer.

\subsection{Rules to Unify Representations Between \texttt{if} Branches}
\label{sec:branchmerge}

One of the main strengths of GL is that it allows case splits that
could otherwise lead to combinatorial blow-ups to be represented
symbolically in Boolean function objects.  As an abstract example,
suppose each function \texttt{f$n$} in the term below produces a
10-bit natural:
\begin{verbatim}
(let* ((x1   (if (c1 x0) (f11 x0) (f12 x0)))
       (x2   (if (c2 x1) (f21 x1) (f22 x1)))
       (x3   (if (c3 x2) (f31 x2) (f32 x2)))
       (x4   (if (c4 x3) (f41 x3) (f42 x3)))
      ...)
    x10)
\end{verbatim}
\noindent Proving this using ACL2 would likely require a case-split
into 1024 cases, but those cases can all be represented simultaneously
in the bits of one symbolic number.  Each \texttt{if}
term produces a number on each branch, which are merged into a common
symbolic representation before continuing on with the next term.
These potential case splits are then handled in the BDD engine or SAT
solver rather than by explicit enumeration.

To extend this feature to term-level representations, GL allows rules
to merge branches of \texttt{if}s that result in different term-level
representations into a unified representation. An
example for the theory of records:
\begin{verbatim}
(gl::def-gl-branch-merge merge-if-of-s
 (equal (if c (s k v rec1) rec2)
        (s k
           (if c v (g k rec2))
           (if c rec1 rec2))))  
\end{verbatim}
\noindent The left-hand side of this rule should always be an \texttt{if} with
variables for the test and the else branch, but some function call for
the \texttt{then} branch.  It is applied to both branches
symmetrically, so in this case if a call of \texttt{s} appears in the
else but not then branch, it will still be applied (it effectively
matches the same \texttt{if} with the test negated and branches
swapped).
This rule is particularly useful when the two branches of an \texttt{if} update
different keys of a record.  When using this rule it is best to also
have rewrite rules in place to eliminate redundant sets of the same key, since
otherwise setting the same key on both branches will cause the set of
the key to be nested.

\subsection{Function Call Shape Specifier}
\label{sec:callshapespec}

Shape specifiers, given in the \texttt{:g-bindings} argument to
\texttt{def-gl-thm}, bind term variables to symbolic representations.
Without the other term-level features discussed here, the only
practical way to use GL was to provide such a binding for each
variable, usually to a symbolic Boolean, a symbolic integer of fixed
bit width, or a cons structure containing symbolic Booleans and/or
integers.  The Boolean variable generation feature discussed in
Section~\ref{sec:vargen} makes it possible to prove some practical
theorems while leaving variables unbound (effectively binding them to
term-level variables named after themselves). But when using
term-level features it can still be advantageous to choose a BDD
variable ordering or otherwise control the representation of theorem
variables.  The \texttt{g-call} shape specifier is a new feature that
allows a theorem variable to be bound to a term-level function call.
The difficulty of using such a construct is in proving that the
shape specifier covers the required range of objects. If we have the
hypothesis \texttt{(signed-byte-p 8 x)}, then it is easy to prove that
an 8-bit integer shape specifier covers all needed inputs, but if the
shape specifier instead is a function call then we need some knowledge
about the function.

The coverage obligation for a given shape specifier is to show that
for any target object satisfying some assumption (the theorem
hypotheses, at the top level), there exists an environment under which
the symbolic object generated from the shape specifier evaluates to
the target object.  For most kinds of shape specifiers, this can be
determined syntactically: as noted above, an 8-bit integer shape
specifier is known to cover all 8-bit integers, because every bit is
required to be an independent Boolean variable.  But to show that a
function applied to some shape specifier arguments covers some object,
we need to know what arguments to that function produce that object,
i.e., we need an inverse function.  The \texttt{g-call} shape
specifier therefore contains not only the function and argument list
of the call, but also a third field giving the inverse function, which
must be a single-argument function symbol or lambda.  Applied to a
target object in the required coverage range, this function should
produce an argument list on which the function produces that target
object.  We can then split the coverage obligation into two parts: the
function applied to its inverse produces the target object, and the
inverse is in the range of the argument shape specifiers.

To relieve the coverage obligation for the shape specifiers, GL generates a 
proof obligation term which it returns from the clause processor as a side goal.  The proof obligation term is computed by the function \texttt{(shape-spec-oblig-term
  $\var{sspec}$ $\var{target}$)}.  When the given shape specifier
$\var{sspec}$ does not contain any \texttt{g-call} objects, the obligation is stated as
 \texttt{(shape-spec-obj-in-range $\var{sspec}$
  $\var{target}$)}, which syntactically determines whether the shape
specifier covers the target.  On a \texttt{g-call} shape specifier, \texttt{shape-spec-oblig-term}
produces two obligations using the provided inverse term.  If $\var{sspec}$ is \texttt{(g-call $\var{fn}$
  $\var{args}$ $\var{inv}$)}, the obligations are:
\begin{enumerate}
\item The function applied to the argument list produced by the
  inverse function on the target object equals the target object (here
  $n$ is the length of $\var{args}$ minus 1):
\begin{lstlisting}
 (let ((inv-args (<$\var{inv}$> <$\var{target}$>)))
   (equal (<$\var{fn}$> (nth 0 inv-args) <$\ldots$> (nth <$n$> inv-args)))
          <$\var{target}$>)
\end{lstlisting}
\item The shape specifiers of the argument list recursively cover the
  values produced by the inverse function when applied to the target
  object: the obligations showing this are generated by recursively
  calling \texttt{shape-spec-oblig-term} on the arguments and the
  corresponding \texttt{nth}s of the call of the inverse function.
\end{enumerate}

Here is an example:
\begin{verbatim}
(defun plus (st dest src1 src2)
  (s dest (loghead 10 (+ (g src1 st) (g src2 st))) st))

(defun s-inverse (key obj)
  (list key (g key obj) obj))

(gl::def-gl-thm plus-c-a-b-correct
  :hyp (and (unsigned-byte-p 9 (g :a st))
            (unsigned-byte-p 9 (g :b st)))
  :concl (let* ((new-st (plus st :c :a :b))
                (a (g :a st))
                (b (g :b st))
                (new-c (g :c new-st)))
           (equal new-c (+ a b)))
  :g-bindings `((st ,(gl::g-call
                      's
                      (list :a
                            (gl::g-int 0 1 10)
                            (gl::g-call
                             's
                             (list :b
                                   (gl::g-int 10 1 10)
                                   (gl::g-var 'rest-of-st))
                             '(lambda (x) (s-inverse ':b x))))
                      '(lambda (x) (s-inverse ':a x)))))
  :cov-theory-add '(nth-const-of-cons
                    s-same-g
                    s-inverse))
\end{verbatim}
The first two arguments in each \texttt{g-call} are the function name
and argument list, and the third argument gives the inverse function.
In the example, the variable \texttt{st} is bound to an object
representing the following term, where $a$ and $b$ are
10-bit integer shape specifiers:
\begin{lstlisting}
(s :a <$a$> (s :b <$b$> rest-of-st))
\end{lstlisting}

The inverse function for each \texttt{g-call}
produces a list of three values, corresponding to the arguments
$\var{key}$, $\var{val}$, $\var{rec}$ of \texttt{s}.  The key is
chosen to match the key used by the call, the value is chosen to match
the existing binding of the key in the record, and the record is just
the input record itself.  By the record property \texttt{s-same-g},
\texttt{(equal (s a (g a r) r) r)}, this reduces to the input record,
satisfying obligation 1.\footnote{The fact that we don't need to
  assume that the input value is a record is due to a special property
  of the records library, which goes to some pains to ensure that its
  functions work together logically without type assumptions.}  The
shape specifiers for the arguments \texttt{(key val rec)} also cover the
values, satisfying Obligation 2: the keys are equal constants; the value shape specifiers are each
10-bit integers, which by the \texttt{unsigned-byte-p} hypotheses
suffice to cover the specified components, and the \texttt{g-var}
construct used as the innermost record can cover any object.  The coverage
proof completes with the help of the \texttt{:cov-theory-add} argument,
which enables the listed rules during the proof of the coverage obligation.

\section{Implementation Details}
\label{sec:details}


While an in-depth description of the implementation of these features
is beyond the scope of this paper, we mention here some implementation
details that may be relevant to users.

\subsection{Priority Order of Function Call Reductions}
\label{sec:fncalls}
As the GL symbolic interpreter crawls over a term, at each function
call subterm it first interprets the arguments to the call, reducing
them to symbolic objects (though possibly term-like ones).  Then there
are several reductions that may be used to resolve the function call
to a symbolic object.  These reductions are affected by the setting of
some functions as uninterpreted, as first discussed in
Section~\ref{sec:rewriting}; these functions are stored in a table
where their names are bound to \texttt{NIL} (interpreted),
\texttt{T} (uninterpreted), or \texttt{:CONCRETE-ONLY} (uninterpreted,
but allowed to be concretely evaluated).  The reductions are listed
here in the order they are considered.\footnote{This is coded in the
  book ``centaur/gl/glcp-templates.lisp'' under \texttt{(defun
    interp-fncall ...)}; this template is expanded to an actual
  function definition when a new GL clause processor is created using
  \texttt{def-gl-clause-processor}.}
\begin{enumerate}
\item If the arguments are all syntactically concrete (implying they
  each have only one possible value), and the function is allowed to
  be concretely evaluated (its uninterpreted setting is not
  \texttt{T}), then the
  function is evaluated on the arguments' values using \href{http://www.cs.utexas.edu/users/moore/acl2/manuals/current/manual/?topic=ACL2____MAGIC-EV-FNCALL}{\underline{\texttt{magic-ev-fncall}}} and the resulting value is returned, properly quoted as a concrete object.
\item Rewrite rules are read from the function's \texttt{lemmas}
  property; any rules whose runes are listed in the
  \texttt{gl-rewrite-rules} table are attempted.  If any such rule is
  successfully applied, the resulting term is symbolically interpreted
  in place of the function call.
\item If the function has a custom symbolic counterpart defined in the
  current clause processor \cite{gl-diss} (see documentation topic
  \href{http://www.cs.utexas.edu/users/moore/acl2/manuals/current/manual/?topic=GL____CUSTOM-SYMBOLIC-COUNTERPARTS}{\underline{gl::custom-symbolic-counterparts}}),
  it is run on the arguments and its value returned.
\item If the function is uninterpreted, a new function call object is
  returned containing the function name applied to the argument
  objects.
\item Otherwise, the body of the function is interpreted with the
  arguments bound to its formals.  The particular definition used may
  be overridden; see documentation topic
  \href{http://www.cs.utexas.edu/users/moore/acl2/manuals/current/manual/?topic=GL____ALTERNATE-DEFINITIONS}{\underline{gl::alternate-definitions}}.
\end{enumerate}

\subsection{Handling of \texttt{if} Terms}

There are two main parts of the handling of \texttt{if} terms:
interpreting the test, and merging the branches when the test is not
resolved to a constant value.

First, the test term is symbolically interpreted, resulting in some
symbolic object.  That symbolic object is coerced to a Boolean
function representation as follows:\footnote{Coded in
  ``centaur/gl/glcp-templates.lisp'' under \texttt{(defun
    simplify-if-test ...)}}
\begin{enumerate}
\item If it is a symbolic Boolean, extract its Boolean function.
\item If it is a symbolic number, return constant true.
\item If it is a cons, return constant true.
\item If it is syntactically concrete, return constant true its value is if non-\texttt{nil}, constant false if \texttt{nil}.
\item If it is an if-then-else, recursively extract the Boolean
  function of the test.  If that function is syntactically constant true or constant false, return
  the Boolean function of the corresponding branch; else get the
  Boolean functions of both branches and return the
  Boolean if-then-else of the test and branches.
\item If it is a variable term, return its associated Boolean variable
  if it has one, else generate a fresh Boolean variable, record its
  association, and return it.
\item If it is a function call term matching \texttt{(not $x$)} or
  \texttt{(equal $x$ nil)}, then recursively extract the Boolean
  function for $x$ and return its negation.
\item If it is a function call term matching
  \texttt{(gl::gl-force-check-fn $x$ $\var{strong}$ $\var{dir}$)},
  recursively extract the Boolean function for $x$ and use
  satisfiability checking to try to reduce it to constant-true and/or
  constant-false; see the documentation of
  \href{http://www.cs.utexas.edu/users/moore/acl2/manuals/current/manual/?topic=GL____GL-FORCE-CHECK}{\underline{gl::gl-force-check}}.
\item If it is a function call term not matching any of the above,
  return its associated Boolean variable if it has one, else generate
  a fresh Boolean variable, record its association, update the
  constraint database, and return the variable.
\end{enumerate}

If the resulting Boolean function is syntactically constant
or known true or false under the path condition, then only
one of the two branches needs to be followed, and no merging is
necessary.  Otherwise, each branch is symbolically interpreted with
the path condition updated to reflect the required value of the test,
and the results are merged.  Merging works as follows:\footnote{Coded
  in ``centaur/gl/glcp-templates.lisp'' under \texttt{(defun
    merge-branches ...)}}
\begin{enumerate}
\item If both branch objects are equal, return that object.
\item If the $\var{then}$ branch is a function call object (or a
  cons object, which is viewed as a call of \texttt{cons} here), try
  to rewrite the \texttt{if} with all branch merge rules stored
  under the function symbol.  If any succeeds, proceed to
  symbolically interpret the resulting term.
\item Similarly if the $\var{else}$ branch is a function call or
  cons object, try to rewrite the inverted \texttt{if} term
  \texttt{(if $\neg\var{test}$ $\var{else}$ $\var{then}$)} with branch
  merge rules.
\item If both branches are calls of the same function or both
  conses, recursively merge the corresponding arguments and
  interpret the application of the function to the merged arguments.
\item Otherwise, merge the two branch objects without further symbolic
  interpretation or rewriting, merging components of similar type and
  making \texttt{if} objects when necessary \cite{gl-diss}.
\end{enumerate}

\section{Record Example}
\label{sec:example}

We'll develop a useful theory here for reasoning about records in GL,
under the assumption that keys will generally be concrete values; this
won't work as well for memories where addresses are sometimes
computed, for example.

To effectively reason about records, first it is important that the
basic functions involved are uninterpreted.  Here we use the
\texttt{:concrete-only} option so that they can be concretely executed
when necessary.
\begin{verbatim}
(gl::gl-set-uninterpreted g :concrete-only)
(gl::gl-set-uninterpreted s :concrete-only)
\end{verbatim}
The most important rule for reasoning about records is this rule,
which determines the value extracted by a lookup after an update:
\begin{verbatim}
(gl::def-gl-rewrite g-of-s-casesplit
  (equal (g k1 (s k2 v x))
         (if (equal k1 k2)
             v
           (g k1 x))))
\end{verbatim}
It is often desirable to normalize the ordering of keys in nests of
\texttt{s} operators, to avoid growing terms out of control.  Here
\texttt{general-concretep} and \texttt{general-concrete-obj} respectively
check whether a symbolic object is constant and return its constant
value.  The third hypothesis of the following rule ensures that the
rule does not loop and normalizes the ordering of keys to \texttt{<<}
sorted order.
\begin{verbatim}
(gl::def-gl-rewrite s-of-s-normalize
  (implies (syntaxp (and (gl::general-concretep k1)
                         (gl::general-concretep k2)
                         (not (<< (gl::general-concrete-obj k1)
                                  (gl::general-concrete-obj k2)))))
           (equal (s k1 v1 (s k2 v2 x))
                  (if (equal k1 k2)
                      (s k1 v1 x)
                    (s k2 v2 (s k1 v1 x))))))
\end{verbatim}
Other disciplines for organizing the record structure may perform
better in certain contexts: it might be preferable to avoid sorting
keys, or to choose a representation that allows a specialized
structure such as a tree or fast alist to be used for record lookups.

To generate good counterexamples for theorems involving \texttt{g} and
\texttt{s}, the following rule shows how to update a record so that
the given key is bound to a particular value:
\begin{verbatim}
(gl::def-glcp-ctrex-rewrite
  ((g k x) v)
  (x (s k v x)))
\end{verbatim}
The following rule shows how to merge branches in which one branch
resulted in an \texttt{s} call:
\begin{verbatim}
(gl::def-gl-branch-merge merge-if-of-s
  (equal (if test (s k v x) y)
         (s k (if test v (g k y)) (if test x y))))
\end{verbatim}
The above rules are generally sufficient to reason about records
provided we only care about equality of certain fields.  If we want to
check equality of records, we need some rewriting tricks.

We use a helper function here that checks that two records are equal
except possibly in their values on a particular list of keys.  This is
equivalent to the quantified function
\begin{verbatim}
(forall k
    (implies (not (member k lst))
             (equal (g k x) (g k y))))
\end{verbatim}
\noindent but the following recursive formulation is easier to reason
about:
\begin{verbatim}
(defun gs-equal-except (lst x y)
  (if (atom lst)
      (equal x y)
    (gs-equal-except (cdr lst)
                     (s (car lst) nil x)
                     (s (car lst) nil y))))

(gl::gl-set-uninterpreted gs-equal-except)
\end{verbatim}
We can use this function to unwind equalities of \texttt{s} operations
to a conjunction of equalities between the stored values for each key
set.  This theorem starts the process (the theorem with swapped LHS
arguments is also needed):
\begin{verbatim}
(gl::def-gl-rewrite equal-of-s
  (equal (equal (s k v x) y)
         (and (equal v (g k y))
              (gs-equal-except (list k) x y))))
\end{verbatim}
This theorem (and the similar theorem with swapped arguments to the LHS
\texttt{gs-equal-except}) unrolls the rest of the way until we reach
the original records at the beginning of the \texttt{s} nest:
\begin{verbatim}
(gl::def-gl-rewrite gs-equal-except-of-s
  (equal (gs-equal-except lst (s k v x) y)
         (if (member k lst)
             (gs-equal-except lst x y)
           (and (equal v (g k y))
                (gs-equal-except (cons k lst) x y)))))
\end{verbatim}
Finally, this theorem accounts for the base case in which we have no
more \texttt{s} calls and just compare the original records directly.
This suffices in the common case where the original record is the same
exact object on both sides of the comparison.
The syntaxp check here ensures that we don't rewrite our
\texttt{equal} hypothesis with \texttt{equal-of-s} and cause a loop.
\begin{verbatim}
(gl::def-gl-rewrite gs-equal-except-of-s-base-case
  (implies (and (syntaxp (and (not (and (consp x)
                                        (eq (gl::tag x) :g-apply)
                                        (eq (gl::g-apply->fn x) 's)))
                              (not (and (consp y)
                                        (eq (gl::tag y) :g-apply)
                                        (eq (gl::g-apply->fn y) 's)))))
                (equal x y))
           (equal (gs-equal-except lst x y) t)))
\end{verbatim}
This rewriting strategy isn't complete for proving equality or
inequality between records.  For example, if we have a comparison of
two different variables and an assumption that they differ on some
key, we don't resolve the equality to false.  However, it does suffice for
the common usage pattern of showing that a machine model
running some program produces the same final state as a specification.

\section{Application}
\label{sec:applications}

The GL features discussed here have been used in proofs about
microcode routines at Centaur Technology \cite{davis2014microcode}.
The proof framework was based on specifying the updates to the machine
state computed by small code blocks, then composing these blocks
together to specify the resulting machine state from a whole microcode
routine.  The correctness proof for each code block would prove that
if the starting state satisfied certain conditions -- such as the
program counter having the correct value, or certain constraints
holding of data elements -- that running the machine model for some
number of steps produced a state equal to that produced by the
specification.  Don't-care fields could be ``wormhole abstracted''
\cite{hardinucode}, essentially specifying their value as whatever was
produced by the implementation.

The machine model used in the microcode proofs was a stobj (for best
concrete execution performance) but was modeled logically as a nested
record structure in which field accessors always fixed the stored
values to the appropriate type to match what was stored in the stobj
(see the documentation for
\href{http://www.cs.utexas.edu/users/moore/acl2/manuals/current/manual/?topic=ACL2____DEFRSTOBJ}{\underline{defrstobj}}).
A GL term-level theory similar to the one described in
Section~\ref{sec:example} was used to handle the accesses and updates,
and Boolean variable generation was used to generate representations
for fields of the state, which were mostly fixed-width natural
numbers.

A code block proof could be attempted either using GL or not; in some
cases the traditional ACL2 proof procedure was preferable.  GL was
successfully used on basic code blocks of up to about 30 instructions,
and also for some block composition theorems.

\section{Soundness Proof}
\label{sec:proof}

The GL clause processor is verified as a sound extension to ACL2.  The
symbolic interpreter at its core is a 22-function mutual recursion,
consisting of four functions responsible for different steps in
interpreting a term, five functions to implement the conditional
rewriter, ten dealing with processing of \texttt{if} terms, two
responsible for adding Boolean variable constraints, and one that
interprets a list of terms.  To reason about this mutual recursion, we
used the utilities from community book ``tools/flag.lisp'' and a
wrapper macro to support proving similar theorems about all the
functions in the mutual recursion.  To get the proofs about this
mutual recursion to perform acceptably, we tried to make each function
as simple as possible, which sometimes involved splitting the
functions into more mutually-recursive parts.

Previous to the addition of the term-level features,
the main theorem about the interpreter was that the evaluation of the
output symbolic object under some environment was equal to the
evaluation of the input term under the alist formed by evaluating each
of the symbolic objects in the input symbolic object bindings.  The 
statement, greatly simplified, looked like this:
\begin{verbatim}
 (implies (and (bool-function-eval pathcond env)
               ...)
          (equal (sym-object-eval (gl-interpret-term x bindings pathcond) env)
                 (term-eval x (sym-object-alist-eval bindings env))))
\end{verbatim}

Of the new features discussed here, the only one that added
significant complexity or difficulty to this theorem was the
generation of fresh Boolean variables (Section \ref{sec:vargen}).
Rewriting and \texttt{if} branch merging rules added functions to the
mutual recursion, but didn't require us to introduce any new concepts
into the proofs.

The Boolean variable generation feature complicates the relationship
between the input bindings and output symbolic object in the theorem
above because the output term may reference Boolean variables that
were created during the course of the symbolic interpretation.  The
meanings of these new Boolean variables are stored in a database that
associates the generated variables (represented as natural numbers,
generated in increasing order) with the symbolic objects they were
generated to represent.  The symbolic interpreter takes a Boolean
variable database as input and returns it, possibly extended with new
variable/object correspondences, as output.  Existing correspondences
in the database are not changed during symbolic interpretation.

In order to prove a correspondence between output symbolic
object and input term and bindings similar to the one above, we need some facts about
the set of Boolean variables on which each object depends:
\begin{itemize}
\item The symbolic object returned by the interpreter may only depend
  on Boolean variables present in the input bindings or bound in the
  Boolean variable database.
\item Symbolic objects bound to Boolean variables in the Boolean
  variable database are strictly ordered in their dependencies: the
  object bound to a given Boolean variable may only depend on
  lower-numbered variables.
\end{itemize}
We also need to assume:
\begin{itemize}
\item The symbolic objects in the input bindings do not depend on any
  Boolean variable numbers greater than the maximum currently present
  in the Boolean variable database.
\end{itemize}
These show that adding a new variable/object correspondence to the
database does not affect the meaning of existing objects.  Finally, an
additional assumption is needed about the environment under which the
output object and input bindings are evaluated:
\begin{itemize}
\item The environment is consistent with the Boolean variable
  database: that is, for each Boolean variable/object binding $(v, o)$
  in the database, the binding of $v$ in the environment must agree
  with the truth value of the evaluation of $o$ under the environment.
\end{itemize}

With these assumptions, we can prove a correspondence between the
output symbolic object and input term and bindings similar to the one
above.  However, we must also show that environments satisfying the
above assumption exist; in particular, given an environment under
which the input shape specifiers evaluate to a particular assignment
of theorem variables, we must show that there exists an environment
that preserves this evaluation and is consistent with the Boolean
variable database.  We do this by iteratively mapping the Boolean
variables from the database to the evaluations of their corresponding
objects, in order from the lowest-numbered.  The result, we prove, is
an environment that preserves the evaluation of the shape specifiers
and is consistent with the database.  This allows us to argue that if
the Boolean function representation of the theorem's conclusion is
valid, then the theorem is true: given an assignment to the theorem's
variables under which the theorem is false:
\begin{enumerate}
\item this assignment satisfies the hypothesis, so by coverage there
  exists an environment under which the shape specifiers evaluate to
  that assignment;
\item we can extend that environment to be consistent with the Boolean
  variable database while preserving the evaluation of the shape
  specifiers;
\item therefore the result of symbolic simulation of the conclusion
  under that environment must agree with the evaluation of the theorem
  under that assignment;
\item but since the symbolic simulation result is a valid Boolean function,
   the theorem is true.
\end{enumerate}

\section{Future Extensions}
\label{sec:future}

\subsection{Fixtypes Support}
\label{sec:fixtypes}
The FTY library (see documentation for
\href{http://www.cs.utexas.edu/users/moore/acl2/manuals/current/manual/?topic=ACL2____FTY}{FTY})
provides automation for defining (mutually-)recursive sum-of-products,
product, list, and alist types that support a fixing discipline that
avoids the need to assume that variables are well-typed when proving
theorems \cite{fixyourtypes}.  In this fixing discipline, all types
have corresponding fixing functions, which logically fix their
argument to the corresponding type but are essentially free to execute
in guard-verified code because the inputs are known to already be of
the correct type.  Unfortunately, this fixing discipline could
potentially have a very bad performance impact in GL, where the
logical version of the code is used and guards are ignored.  This also
affects performance when calling discipline-compliant functions on
concrete values, because each call from the symbolic interpreter must
first check its guard.

For cases where the fixing discipline becomes a practical problem, it
may be a better approach to treat the types involved as abstract
datatypes.  The FTY library already provides rewrite rules for the
ACL2 rewriter that do this, and some of these could be used directly in
GL.  Other rules
necessary for GL to effectively reason about an FTY type could be
generated automatically.  The FTY
type definition events leave behind information in ACL2 tables that
fully describe the types.  It would be reasonable to use this
information to generate a GL term-level theory for a given FTY type.

\subsection{Feature Parity with the ACL2 Rewriter}
\label{sec:parity}
GL's rewriter is missing several important features of ACL2's
simplifier.  We list these in order from most to least likely to be
worth the implementation effort.
\begin{itemize}
\item Bind-free and binding hypotheses are useful for effective
  programming of the rewriter and would be relatively easy to
  implement.  The primary burden would be to modify the correctness
  proof of the rewriter/symbolic interpreter; we would need to show that these
  hypotheses only extend the current unifying substitution with free variables, and
  that such extensions to the substitution are allowable when applying a
  rewrite rule.

\item Congruence-based reasoning is also somewhat likely to be useful
  for programming the rewriter, and is partly implemented already.  GL
  already distinguishes between \texttt{equal} and \texttt{iff}
  contexts and the rewriter variable that makes this distinction
  already has been extended to handle other equivalences; the
  correctness proof also accounts for arbitrary equivalences.  The
  parts yet to be implemented are the parsing and checking of
  congruence, refinement, and equivalence rules so that we can soundly
  use these theorems, and the application of congruences to the
  equivalence context when beginning to interpret a new subterm.

\item Free-variable matching hypotheses are somewhat of a mismatch to
  GL's reasoning paradigm.  Whereas ACL2 has a clause from which it
  derives a type-alist that stores its assumptions, GL only has a path
  condition.  The path condition typically stores assumptions that are
  Boolean formulas rather than terms that are known true.
  Nevertheless, we could scan the path condition for known-true terms
  that unify with the current hypothesis under the current
  substitution.  The proof obligations would be the same as for the
  bind-free and binding hypotheses, so only this implementation work
  would be necessary.

\item Forward-chaining is an ACL2 feature that also doesn't fit
  cleanly into GL's paradigm, for similar reasons as for free-variable
  matching hypotheses.  We could allow forward reasoning when adding
  an \texttt{if} test to the path condition, which would require some
  non-insurmountable amount of new implementation work.  The Boolean
  variable constraint system (Section \ref{sec:constraintgen}) already
  allows forward reasoning after a fashion, but this doesn't affect
  rewriting directly since the constraints generated are currently
  only used in Boolean proofs.
\end{itemize}

\subsection{Debugging Support}

GL currently does not have any support beyond ACL2's \texttt{trace\$}
utility for debugging failed term-level proofs.  While
\texttt{trace\$} is extremely useful to a user knowledgeable enough
about implementation details to know what functions to trace,
targeted debugging tools such as ACL2's 
\href{http://www.cs.utexas.edu/users/moore/acl2/manuals/current/manual/?topic=ACL2____BREAK-REWRITE}{break-rewrite}
would be a dramatic usability improvement.

\section{Conclusion}
\label{sec:conclusion}

This paper describes extensions to the GL bitblasting framework that
allow it to deal in abstract terms, rather than a limited value-like
representation.  The GL framework is available under a permissive
license in the ACL2 community books.

\section*{Acknowledgements}
We would like to acknowledge Jared Davis for helpful conversations
during the implementation of the features discussed, and Shilpi Goel
for testing out some of these features during her dissertation
research.  We also thank the reviewers for helpful comments.

\bibliographystyle{eptcs}
\bibliography{paper}
\end{document}